\documentclass[MSNbibl,number,citesort,dvips]{arxbj}
\usepackage{graphicx}
\aid{0}
\volume{19}
\issue{5A}
\pubyear{2013}
\firstpage{1920}
\lastpage{1937}
\doi{10.3150/12-BEJ435} %kopijuoti is PTS

\makeatletter
\renewcommand{\emptyset}{\varnothing}
\newcommand{\diag}{\operatorname{diag}}
\newcommand{\rank}{\operatorname{rank}}

\newproclaim{definition}{Definition}
\newremark{example}{Example}
\newtheorem{theo}{Theorem}
\newtheorem{prop}{Proposition}
\newremark{rem}{Remark}
\newtheorem{lemma}{Lemma}
\makeatother

\begin{document}
\begin{frontmatter}

\title{Identification of discrete concentration graph models
with one hidden binary variable}
\runtitle{Identification of discrete graphical models with hidden nodes}

\begin{aug}
%%%% inicialai - be tarpu
\author[1]{\fnms{Elena} \snm{Stanghellini}\corref{}\thanksref{1}\ead[label=e1]{elena.stanghellini@stat.unipg.it}} \and
\author[2]{\fnms{Barbara} \snm{Vantaggi}\thanksref{2}\ead[label=e2]{vantaggi@dmmm.uniroma1.it}}
\runauthor{E. Stanghellini and B. Vantaggi}
\address[1]{D.E.F.S., Universit\`a di Perugia, Perugia, Italy. \printead{e1}}
\address[2]{S.B.A.I., Universit\`a ``La Sapienza'', Roma, Italy. \printead{e2}}
\end{aug}

% HISTORY:
\received{\smonth{3} \syear{2011}}
\revised{\smonth{2} \syear{2012}}

% ABSTRACT
%
\begin{abstract}
Conditions are presented for different types of identifiability of
discrete variable models generated over an
undirected graph in which one node represents a binary hidden variable.
%for identifiability of discrete
%undirected graphical models with a binary hidden node.
These
models can be seen as extensions of the latent class model to
allow for conditional associations between the observable random variables.
Since local identification corresponds to full rank of the
parametrization map, we establish a necessary and sufficient
condition for the rank to be full everywhere in the parameter
space. The condition is based on the topology of the undirected
graph associated to the model. For non-full rank models, the obtained
characterization allows us to find the subset of the parameter space
where the identifiability breaks down.
\end{abstract}

% KEYWORDS
%
\begin{keyword}
\kwd{conditional independence}
\kwd{contingency tables}
\kwd{finite mixtures}
\kwd{hidden variables}
\kwd{identifiability}
\kwd{latent class}
\kwd{latent structure}
\kwd{log linear models}
\end{keyword}

\end{frontmatter}

%s1 #&#
\section{Introduction}\label{s1}

Statistical models with latent variables have become important tools in
applied studies, as they allow to
include the effects of unobservable variables over the observable ones
and to
correct for the possible distortion induced by heterogeneity in the data.
However, it is now widely recognized that when some of the variables
are never observed,
standard statistical procedures may be problematic, as
non-identifiability of the parameters and local maxima in the
likelihood function can
occur.

In this paper, we focus on local identifiability of undirected
graphical models for discrete variables with one binary hidden, or
latent, variable. Note that models with a binary latent variable
arise in several studies, as those concerning the absence/presence
of a particular trait. In a recent paper by Allman \textit{et al.} \cite{Allmanetal},
a weaker form than local identification has been treated and named
generic identification in which case a set of non-identifiable
parameters may be present which resides in a subset of null
measure.
%Their derivations however,
%do not allow to establish the explicit expression
%of the subspace of null measure where identifiability breaks down.
To find the explicit expression of such subset is important, since
standard statistical procedures may
fail if the estimates of the parameters are close to the singular
locus; see, for example, \cite{Drton}.

Since, by the inverse function theorem, local identifiability
corresponds to
%Since local identification corresponds to
full rank of the parametrization map, we establish a necessary and
sufficient condition for the rank to be full everywhere in the
parameter space. The condition is based on the topology of the
undirected graph associated to the model. This contribution is
similar to what is done in \cite{Drtonetal} for linear
structural equation models. For non-full rank models, the obtained
characterization allows us to find the subset where the
identifiability breaks down.

In Section \ref{s2}, the class of models is presented together with the
notion of identification. The main theorem is in Section \ref{s3}. In
Section \ref{s4}, we present the derivations that lead to the main result.
Section \ref{s5} contains concluding remarks.

%s2 #&#
\section{Discrete undirected graphical model}\label{s2}

Let $G^K=(K,E)$ be an undirected graph with node set
$K=\{0,1,\ldots,n\}$ and edge set $E=\{(i,j)\}$ whenever vertices
$i$ and $j$ are adjacent in $G^K$, $0\leq i< j \leq n$. To each
node, $v$ is associated a discrete random variable $A_v$ with finitely
many levels. A
discrete undirected graphical model is a family of joint
distributions of the variables $A_v$, $v \in K$, satisfying the
Markov property with respect to $G^K$, namely that the joint
distribution of the random variables
factorizes according
to $G^K$; see
\cite{Lauritzen}, Chapter 3, for definitions and concepts.

\begin{figure}[b]

\includegraphics{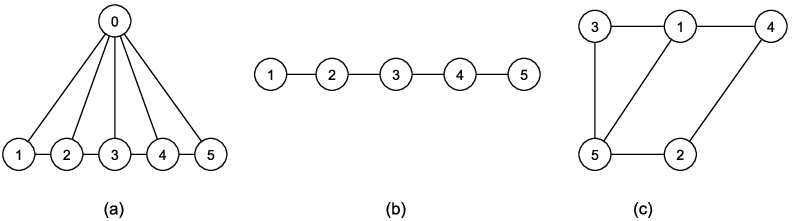}

\caption{Example of (a) a $G^K$ graph and the corresponding graphs
(b) $G^O$ and (c) $\bar G^O$.} \label{prima}
\vspace*{-9pt}\end{figure}

Let $A_0$ be a binary latent variable and $O=\{1,\ldots,n\}$
be the set of nodes associated to observable random variables. In the
following, let $G^B$ be the (sub)graph
$G^B=(B,E_B)$ of $G^K$ induced by $B\subseteq K$. We denote with
$\bar G^B=(B, \bar E_B)$ the complementary graph of the (sub)graph
$G^B$, where $\bar E_B$ is the edge set formed by the pairs
$(i,j)\notin E_B$ with $i,j\in B$ ($i\neq j$). In Figure~\ref{prima}(b) and (c), the graph $G^O$ and its complementary graph
$\bar
G^O$ associated to the graph $G^K$ of Figure~\ref{prima}(a) is
presented.

%f1 #&#

Let $l_v$ denote the number of levels of $A_v$, $v\in K$, and let
$l=\prod_{v=1}^n l_v$. Without loss of generality, we assume that
the variable $A_v$ takes value in $\{0,\ldots,l_v-1\}$. We
consider the multidimensional contingency table obtained by the
cross classification of $N$ objects according to $A_v$. Let $X$ be
the $2l \times1$ vector of entries of the contingency table,
stacked in a way that the levels of $A_0$ are changing slowest.

Data for contingency tables can be collected under various
sampling schemes; see \cite{Lauritzen}, Chapter~4. We assume for now
that the elements of $X$ are independent Poisson random variables
with $E(X)=\mu_X$.

Let $\log\mu_X= Z \beta$, where $\beta$ is a $p$-dimensional
vector of unknown parameters;
$Z$ is
a $2l \times p$ design matrix defined in a way that the joint
distribution of $A_v$, $v \in K$, factorizes according to $G^K$
and such that the model is graphical. We assume $\beta\neq0$ and let
$\Omega$
be parameter space, $\Omega=(\mathbb{R}\setminus0)^p$. This implies
that for each complete subgraph $G^S=(S,E_S)$, $S\subseteq K$,
there is a non-zero interaction term of order $|S|$ among the variables
$A_v$, $v \in S$.
%$Z$ is a $2l \times p$ design matrix defined in a way that the joint
%distribution of $A_v$, $v \in K$, factorizes according to $G^K$
%and such that the model is graphical. We assume $\beta\neq0$ and let
%$\Omega$
%be parameter space $\Omega=\mathbb{R}^p\setminus0$

We adopt the corner point
parametrization that takes as first level the cell with $A_v=0$,
for all $v \in K$, see, for example, \cite{Darroch}. We denote by $Y$ the
$l \times1$ vector of the counts in the marginal table, obtained
by the cross classification of the $N$ objects according to the
observed variables only. The vector $Y$ is stacked in a way that
$Y=LX$, with $L=(1,1) \otimes e_{l}$, where $e_l$ is the identity
matrix of dimension $l$. By construction, the
elements of $Y$ are independent Poisson random variables with
$\mu_Y=L\mathrm{e}^{Z\beta}$.

If we denote with $\psi$ %: \mu_Y \rightarrow\beta$
the parametrization map from the natural parameters $\mu_Y$ to the new
parameters $\beta$,
global identifiability, also known as as strict identifiability,
corresponds to injectivity of $\psi$, while, when $\psi$ is
polynomial, local identifiability corresponds to $\psi$ being
finite-to-one. As argued in \cite{Allmanetal}, there may
be models such that the parametrization mapping is finite-to-one
almost everywhere (i.e., everywhere except in a subset of null
measure). In this case, we speak of generically
identifiable models.

By the inverse function theorem, a model is locally
identified if the rank of the transformation from the natural parameters
$\mu_Y$ to the new parameters $\beta$ is full everywhere in the
parameter space $\Omega$.
This is
equivalent to the rank of the following derivative matrix\looseness=1
%
%e1 #&#
\begin{equation}\label{erre}
D(\beta)^T=\frac{\partial\mu_Y^{T}}{\partial\beta}= \frac
{\partial
(L \mathrm{e}^{Z\beta})^{T}}{\partial\beta}=(LRZ)^{T}
\end{equation}\looseness=0
being full, where $R=\diag(\mu_X)$. Note that the
$(i,j)$th element of $D(\beta)$ is the partial derivative of the
$i$th component of $\mu_Y$ with respect to $\beta_j$ the $j$th
element of $\beta$.

The multinomial case can be addressed in an analogous way to the
Poisson, after noting that the rank of the matrix $D(\beta)$ and the
rank of
its submatrix $D_0(\beta)$ obtained by
deleting the last column are the same.

Note that, by setting $t_j=\mathrm{e}^{\beta_j}$ for any parameter
$\beta_j$, the parametrization map turns into a polynomial
one. This implies, see, for example, \cite{Pachter}, Chapter 1, that if there
exists a
point in the parameter space of $t_j$, and therefore in $\Omega$,
at which the Jacobian has full rank, then the rank is full almost
everywhere. Therefore, either there is no point in the parameter
space at which the rank is full, or the rank is not full in a
subset of null measure. The object of this paper is (a)~to
establish a necessary and sufficient condition for the rank of
$D(\beta)$ to be full everywhere and (b)~to provide expressions of
the subset of null measure where identifiability breaks down.

%s3 #&#
\section{Main results}\label{s3}

The following definition introduces a graphical notion that is
recalled in the main theorem.

%de1 #&#
\begin{definition}[(Generalized identifying sequence
for a clique)]\label{defini2} A generalized identifying sequence for
a clique $C_0$ of $G^O$ with $|C_0|>1$ is a sequence $\{ S_s\}
_{s=0}^{q}$ in $G^O$ of complete subgraphs such that:
\begin{enumerate}[(b)]
\item[(a)] for $s\in\{0,\ldots, q-1\}$ and for all $i \in
S_s$ there exists a $j\in S_{s+1}$ such that $(i,j)\in\bar E$;
\item[(b)] $|S_{s+1}|\leq|S_{s}|$ for $s\in
\{0,\ldots,q-1\}$, $S_0=C_0$ and $|S_{q}|=1$.
\end{enumerate}
\end{definition}

%ex1 #&#
\begin{example} \label{ex-es2}
Consider the model with graphs $G^K$, $G^O$ and $\bar G^O$ as in
Figure {\ref{prima}}(a)--(c). The clique of $G^O$ are
$\{1,2\}$, $\{2,3\}$, $\{3,4\}$, $\{4,5\}$. For any clique, there is a
generalized
identifying sequence. For $C_0=\{1,2\}$, $S_1=\{4\}$ satisfies the
assumptions of Definition \ref{defini2}.
For $C_0=\{2,3\}$, $S_1=\{5\}$ satisfies the same assumptions.
The same holds for $C_0=\{3,4\}$ and $C_0=\{4,5\}$,
since for both $S_1=\{1\}$ is the required set.
\end{example}

The following theorem characterizes discrete
concentration graph models with one unobserved binary node that
are locally identified everywhere in the parameter space $\Omega$.
The proof is in Appendix \hyperref[AppB]{B}, and uses the results for binary models
developed in Section~\ref{s4}.

%th1 #&#
\begin{theo} \label{teo-ex2}
Let $\beta$ be the vector of parameters of an undirected graphical
model $G^K$ over the discrete
variables $(A_0,A_1,\ldots,A_n)$, with $A_0$ latent binary
variable. Suppose that $(0,u)\notin E$, for any $u\in
T_1\subseteq(K\setminus\{ 0\})$, and $(0,u)\in E$, for all $u \in
S=K\setminus\{0 \cup T_1$\}. A~necessary and sufficient condition
for $D(\beta)$ to be full rank everywhere in the parameter space
is that:
\begin{enumerate}[(ii)]
\item[(i)] $\bar G^S$ contains at least one $m$-clique $C$,
with $m \geq3$;
\item[(ii)] for each clique $C_0$ in
$G^S$ with $|C_0| >1$ there exists a generalized identifying sequence
$S_s$ with all $S_s \subseteq S$.
\end{enumerate}
\end{theo}

The graphical model over the concentration graph as in
Figure {\ref{prima}}(a) is locally identified everywhere in the
parameter space, as condition (i) and (ii) of Theorem
\ref{teo-ex2} are satisfied. This can be checked by noting that the
corresponding $\bar G^S$, $S=O$, contains the 3-clique $\{1,3,5\}$ and
for each clique $C_0$ in $G^O$, $\vert C_0\vert>1$, there is a
generalized identifying sequence as shown in Example \ref{ex-es2}.
%
%f2 #&#
\begin{figure}

\includegraphics{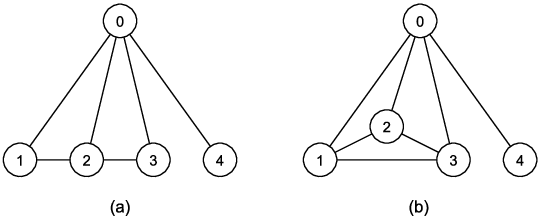}

\caption{Two examples of $G^K$ graphs corresponding to (a) an
identified and (b) a not identified model.} \label{esstanghe}
\end{figure}

Violation of assumption (i) of Theorem \ref{teo-ex2} implies
that $G^S$ either is composed by two and only two complete
components that are not connected or is composed by one connected
component. In the first case, a graphical model is not even
generically identified, i.e. there is no point in the parameter
space such that the parametrization map is full-rank. To see this
let $T_1$ be as in Theorem \ref{teo-ex2} and pose first
$T_1=\emptyset$. Since every clique of $G^S$ corresponds to a
saturated model over the distribution of the observable random
variables conditionally on the latent one, the model is
observationally equivalent to a binary latent class model with two
observable random variables $X^*_j$, $j \in\{1,2\}$, constructed
by clumping the variables in the clique $j$ into\vspace*{1.5pt} a single one.
From \cite{gilula}, without further assumptions, the model is then
rank deficient everywhere in the parameter space. Extension to
$T_1 \neq\emptyset$ follows by noting that the above
considerations hold conditionally on the variables in $T_1$. The
model associated to Figure \ref{esstanghe}(b) is an example.

All other instances of violation of the assumptions of Theorem
\ref{teo-ex2} lead to models that are locally identified almost
everywhere (see Section \ref{s4} and Appendix \hyperref[AppA]{A}). The next example shows
an instance of model which is locally identified almost everywhere
as condition (ii) of Theorem \ref{teo-ex2} fails. The subset
where identifiability breaks down is also presented. It can be
determined throughout the derivations in Section \ref{s4}.

\begin{figure}

\includegraphics{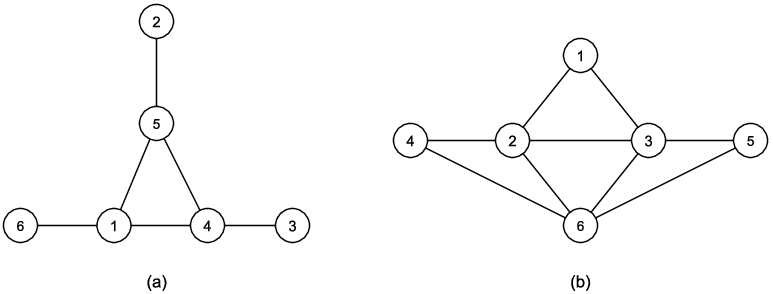}

\caption{The graphs (a) $G^O$ and (b) $\bar G^O$ corresponding to a
model with $S=O$ locally identified almost everywhere.}
\label{seconda}
\end{figure}

%ex2 #&#
\begin{example} \label{es-forcina} With reference to model associated
to the graph in Figure
\ref{seconda}, let $S=O$. The $\bar G^O$ graph contains at least one
3-clique, for example,
$\{1,2,3\}$. The clique
$C_0=\{2,5\}$ has $S_1=\{6\}$ as generalized identifying sequence,
and therefore the corresponding interaction term does not generate
non-identifiability in the parameter space. For symmetry also
$C_0=\{1,6\}$ and $C_0=\{3,4\}$. For $C_0=\{1,4,5\}$, however,
there is no identifying sequence, since $bd_{\bar
G^O}(C_0)=\{2,3,6\}$ is complete in $\bar G^O$.
The subset where identifiability breaks down can be determined from
(\ref{con}) in Appendix \hyperref[AppA]{A},
which also makes clear that it is a subspace. For the binary case:
\[
\cases{
\beta_{\{0,2\}}+\beta_{\{0,2,5\}}=0, \cr
\beta_{\{0,3\}}+\beta_{\{0,3,4\}}=0, \cr
\beta_{\{0,6\}}+\beta_{\{0,1,6\}}=0.}
\]
The rank of $D(\beta)$ is equal to 28 everywhere except
in the above subspace, where it becomes equal to 27.
\end{example}

%f3 #&#

When maximum likelihood estimates are close to the subspace where
identifiability fails, standard asymptotic results may no longer
hold. As an instance, tools for model selection, such as
likelihood ratio test, may be inappropriate, see \cite{Drton}.
Notice further that the model with corresponding $G^O$ graph
obtained by adding the edge $(2,6)$ in Figure~\ref{seconda}(a) is
locally identified everywhere in $\Omega$. Therefore, models
obtained by deleting edges between the observed variables of
locally identified models may not be locally identified everywhere
in the parameter space.

%s4 #&#
\section{Local identification with one binary variable}\label{s4}

In this section, we consider graphical models such that all $n$
observed variables are connected to the
latent one, that is, $(u,0)\in E$, for any observed variable $u
\in O$. We first focus on binary variables only. The assumption will be
relaxed in Theorem \ref{teo-ex1}.
Consider $I\subseteq O$, let $\mu_I$ be the element of $\mu_Y$
associated to the entry of the contingency table having 1 for all
variables in $I$ and 0 for the others. Let $d_I$ be the
row of the matrix $D(\beta)$ corresponding to the first order
partial derivative of $\mu_{I}$ with respect of $\beta$. Note that
$\beta_v$, $v \in K$, represents the main effect of the random
variable $A_v$ and for each subset $I \subseteq O$ such that
$|I|>1$, $\beta_I$ is the interaction term between the variables
in $I$. With $\beta_{\{0,I\}}$, we denote the interaction term
between the variables in $\{0, I\}$. Moreover,
$\beta_{\emptyset}=\mu$ is the general mean. With reference to the
model with concentration graph as in Figure
\ref{esstanghe}(a), let $I=\{1,2\}$. Then $\mu_I$ is the
expected value of the ordered entry $(1,1,0,0)$, $d_I$ is the row
of $D(\beta)$ corresponding to the partial derivative of $\mu_I$
with respect to $\beta$ and $\beta_I$ is the term expressing the
second order interaction between $A_1$ and $A_2$.

With this notation, to each generic $i$-row of $D(\beta)$, we
can associate the set $I$, $I \subseteq O$, of the observed
variables taking value one in row $i$. Each generic column $j$
corresponds to the partial derivatives of $\mu_Y$ with respect to an
element of $\beta$, which we denote with $\beta_J$.
Note that both $I$ and $J$ could
be the empty set. It is then easy to see that if $J\not\subseteq I$,
the generic
$ij$-element of $D(\beta)$ is~0. If $J\subseteq I$,
the $ij$-element of $D(\beta)$ is equal to $\mathrm{e}^{Z_i\beta}$ when
$0\in J$ and to $\mathrm{e}^{Z_i\beta}+\mathrm{e}^{Z_{i+l}\beta}$ otherwise, where
$Z_r$ be the
$r$th row of $Z$.

Furthermore, let $S$ be a complete subgraph of $G^O$ and
$S' \supset S$. For $d_S$ and $d_{S'}$ and $\beta_S$ and
$\beta_{\{0, S\}}$ the $2 \times2$ square sub-matrix of $D(\beta)$
has the following structure:
%
%e2 #&#
\begin{equation}\label{primosistema}
\left[\matrix{
\mathrm{e}^{a}(1+ \mathrm{e}^{b}) &
\mathrm{e}^{a+b} \cr
\mathrm{e}^{a+a'}(1+\mathrm{e}^{b+b'})
&\mathrm{e}^{a+a'+b+b'}}
\right]
\end{equation}
with
\[
a=\mu+\sum_{I \subseteq S}\beta_I,
\qquad b=\beta_0+\sum_{I\subseteq
S}\beta_{\{0, I\}},
\qquad a'=\sum_{\{I \subseteq S', I \not\subseteq
S\}} \delta(I)\beta_{I}
\]
and
\[
b'=\sum_{\{I \subseteq S', I \not
\subseteq S\}} \delta(I)\beta_{\{0, I\}},
\]
 where $\delta(I)=1$ if
$I$ is complete on $G^O$ and 0 otherwise. Matrix
(\ref{primosistema}) is not full rank if and only if $b'=0$.

We first consider the binary latent class model, that is, a model
such that the joint distribution of the random variables factorizes as
follows: $\prod_{v\in O} P(A_v \mid A_0) P(A_0)$, see
\cite{McHugh,Goodman}. From the
assumption $\beta\neq0$, no further independencies than the ones
implied by the above factorization are encoded in the binary latent
class model.

Note that two models with a relabelling of the latent classes,
together with a change of the sign of the $\beta_{\{0,i\}}$,
generate the same marginal distribution over the observable
variables. This issue is known as ``label swapping''.

%pr1 #&#
\begin{prop}\label{prop1} A binary latent class model
is strictly identifiable, up to label swapping, if and only if $n
\geq3$.
\end{prop}

\begin{pf} Sufficiency follows (a) for $n=3$ from
\cite{Allmanetal}, Corollary 2; (b) for $n >3$ from the assumption
$\beta\neq0$. Necessity follows by the fact that if $n
<3$ the model has more parameters than information in the marginal
distribution of the observable random variables.
\end{pf}
%

%$M_i$ in Theorem 1 of \cite{Allmanetal} is a $2 \times2$
%matrix. If a marginal independence between the latent and the
%$j$-th observed variable holds, the corresponding Markov matrix
%$M_j$ has rank one.
%Note that the graph $G^K$ related to a latent class model with $n
%contains a clique of dimension at least 3. In this case $\bar G^O$
%is complete and therefore connected.

We now remove the assumption that the observable random variables are
independent conditionally on the latent one to include a more
general class of graphical models $G^K$ over the variables $A_v$,
$v\in K$. We first
consider graphical models such that $(0,u) \in E$ for all $u \in O$ and
the complementary graphs $\bar
G^O$ are connected and have at least an $m$-clique $C$ with $m
\geq3$.

%pr2 #&#
\begin{prop}\label{pippo1} Let $G^K$ be an undirected graphical model
over the binary variables
$(A_0,A_1,\ldots,\allowbreak A_n)$ with $A_0$ latent and with $(0,u)\in E$,
for all $u \in O$. Assume that in $\bar G^O$ there exists an
$m$-clique $C$, $m \geq3$.
Let $\bar C=\{O \setminus C\}$ and $M_1$ be the
sub-matrix of $D(\beta)$ formed by the rows $d_i$ and $d_{\{i,
j\}}$, with $i \in\bar C$ and $j$ such that $(i,j)\in{\bar E}$,
and by the columns $\beta_{i}$ and $\beta_{\{0,i\}}$. Then $M_1$ has
rank equal to $2|{\bar C}|$ everywhere in the parameter space if and
only if $\bar G^O$ is connected.
\end{prop}

\begin{pf}
If $\bar G^O$ is connected, there exists an ordering (see the
algorithm in Appendix \hyperref[AppA]{A}) of the nodes of $\bar C$ such that for
any $i$, $1\leq i < |\bar C|$, the node $j=i+1$ is such that
$(i,j)\in\bar E$; for $i=|\bar C|$, $j \in C$. Such ordering
generates $|\bar C|$ distinct pairs $(i,i+1)$. Let $M_1^*$ be the
sub-matrix of $M_1$ made up of the rows $d_i$, $d_{\{i, i+1\}}$.
Then $M_1^*$ is a $2|\bar C|$-square lower-block triangular matrix
with blocks $M^i$ associated to row $d_i$, $d_{\{i, i+1\}}$, and
columns $\beta_{i}$ and $\beta_{\{0,i\}}$. The structure of $M^i$
is as (\ref{primosistema}) with $a=\mu+\beta_i$,
$b=\beta_0+\beta_{\{0,i\}}$, $a'=\beta_{j}$ and $b'=\beta_{\{0,
j\}}$ since by construction $(i,j)\in\bar E$. As $\beta_{\{0,
j\}}\neq0$ by assumption, it follows that
$M_1^*$ is full\vspace*{1.5pt} rank and so is $M_1$.

Conversely, if $\bar G^O$ is not connected, then $\bar G^O$
has two or more connected components. Let $\bar G^1=(V_1,E_1)$ and
$\bar G^2=(V_2,E_2)$ be two of them. Consider
any pair of complete sets $I_1\subseteq V_1$ and $I_2\subseteq V_2$
(they could be a singleton) in $G^O$. Note that
$(u,j)\in E$ for any $u \in I_1$ and $j \in I_2$. Therefore, $I_1\cup
I_2$ is a complete subset in $G^O$. Let $S=I_1$ and
$S'$ be any (complete) subset of $I_1 \cup I_2$ such that $S \subset S'$.
From (\ref{primosistema}), any matrix formed by the row $d_S$ and a
row $d_{S'}$, with $S'$
as above, and by the columns $\beta_S$ and $\beta_{\{0,S\}}$ is not full-rank
for $\beta$ such that
%
%e3 #&#
\begin{equation}\label{pippo}
\sum_{\{I \subseteq S', I \not
\subseteq S\}} \beta_{\{0, I\}}=0.
\end{equation}
Then, the submatrix of $M_1$ containing the row $d_S$ and all the
above rows $d_{S'}$ is not full column rank for the above $\beta$,
so $M_1$ is also not full rank.
\end{pf}

Let $t$ be the maximum order of the non-zero interaction terms among
the variables in $O$. For each order $k$, $k \in\{2,\ldots, t\}$,
of interaction between the observable random variables, let $s_k$ be the
number of interaction terms of order $k$. We use $I_{k,r}$ to denote
the set of vertices in $O$ having a non-zero $r$th interaction term
of order $k$, $r \in\{1,\ldots,s_k\}$. Note that, by construction,
$|I_{k,r}|>1$. The following example clarifies the notation.

%ex3 #&#
\begin{example}\label{esempio1} The model with graph $G^K$ as in Figure
\ref{esstanghe}(a) has maximum order $t=2$ and $s_2=2$ with
$I_{2,1}=\{1,2\}$, $ I_{2,2}=\{2,3\}$. The model with graph $G^K$ as
in Figure \ref{esstanghe}(b) has maximum order $t=3$. For $k=2$,
$s_2=3$ with $I_{2,1}=\{1,2\}$, $ I_{2,2}=\{2,3\}$ and
$I_{2,3}=\{1,3\}$; for $k=3$, $s_3=1$ with $I_{3,1}=\{1,2,3\}$.
Similarly, the graph $G^O$ as in Figure \ref{seconda}(a)
has maximum order $t=3$. For $k=2$, $s_k=6$, with $I_{2,1}=\{1,4\}$,
$I_{2,2}=\{1,5\}$, $I_{2,3}=\{1,6\}$, $I_{2,4}=\{2,5\}$,
$I_{2,5}=\{3,4\}$, $I_{2,6}=\{4,5\}$; for $k=3$, $s_k=1$, with
$I_{3,1}=\{1,4,5\}$.
\end{example}

The graphical notion of identifying sequence will be used to
characterize the subset where identifiability breaks down.

%de2 #&#
\begin{definition}[(Identifying
sequence for a complete subgraph)] \label{id-seq} An identifying sequence for a
complete subgraph
$I_{k,r}$ of $G^O$ (with $k\geq2$) is a sequence $\{
I_s\}_{s=0}^{q'+1}$ of complete subgraphs, $q'\geq0$, of $G^O$
such that $I_0=I_{k,r}$, $I_s\neq I_{s^{\prime}}$
(for $s\neq s^{\prime}$) with $s,s^{\prime} \in\{0,\ldots,q'+1\}$ and
satisfying the following assumptions:
\begin{enumerate}[(b)]
\item[(a)] for all $s \in\{0,\ldots,q'\} $ and for all $i\in I_s $ there
exists a $j\in I_{s+1}$ such that $ (i,j)\notin E$;
\item[(b)] for all $s \in\{0,\ldots,q'\}$, $
|I_s|=k $ and $|I_{q'+1}|<k$.
\end{enumerate}
\end{definition}

An equivalent formulation of condition (a) is that for $s \in\{
0,\ldots,q'\}$ and for all $i\in I_{s}$ there
exists a $j\in I_{s+1}$ such that $i$ and $j$ are connected in the
complementary graph $\bar G^O$.

%re1 #&#
\begin{rem}\label{rem-1} If there exists a sequence of complete subgraphs
satisfying (a), but such that $|I_s|>k$, for some $s \in\{1,\ldots
,q'\}$,
then there exists also a sequence satisfying $|I_s|=k$:
as a matter of fact, if for all $i\in I_s$ there exists a node
$j\in I_{s+1}$ such that $(i,j)\notin E$, then $I_{s+1}$ can be
chosen in a way that $|I_{s+1}|$ cannot be greater than $|I_{s}|$.
Therefore, if a complete subgraph $I_{k,r}$ admits no identifying
sequence of complete subgraphs, then either there is no sequence
of $I_s$ such that (a) is satisfied or there is no $I_{q'+1}$
such that $|I_{q'+1}|<k$.
\end{rem}

%re2 #&#
\begin{rem}\label{rem0} For any identifying sequence $\{
I_s\}_{s=0}^{q'+1}$ related to a complete subgraph $G^O$, $I_s\cap
I_{s+1}=\emptyset$ holds, as if, by absurd, $i\in I_s\cap I_{s+1}$, then
$(i,k)\in E$ for any $k\in I_{s+1}$ (since $I_{s+1}$ is complete in
$G^O$), which contradicts the assumptions.
\end{rem}

Given an identifying sequence $\{ I_s\}_{s=0}^{q'+1}$, related to a
complete set $I_{k,r}$, let $V\subseteq I_{s+1}$ and
\[
I_{s}^V=\bigcap_{j\in V}\{i\in I_{s}\dvtx(i,j)\in E\}
\]
be
the subset of $I_s$ with nodes connected in $G^O$ to any node $j$
belonging to $V$. Note that, from Remark \ref{rem0}, for
$V=I_{s+1}$, $I_{s}^{V}=\emptyset$.

%re3 #&#
\begin{rem}\label{rem3} If there is an identifying sequence
satisfying (a) but such that $I_s=I_{s^{\prime}}$ for some $s\neq
s^{\prime}, s<s^{\prime}$, then there is also a shorter identifying
sequence, which is constructed by excluding
the interactions from $I_{s+1}, \ldots,I_{s^{\prime}}$.
\end{rem}

%re4 #&#
\begin{rem} \label{rem5}
The fact that the assumptions (a)--(b) hold for all complete
subgraphs $I_{k,r}$ does not imply that they hold also for the all
complete subgraphs $I_{k',v}$ such that $I_{k',v}\supset I_{k,r}$.
The graph in Figure \ref{seconda}(a) is an example, as for
each complete subgraph of $G^O$ such that $k=2$ there is an identifying
sequence. However,
there is no identifying sequence for $I_{3,1}=C_0=\{1,4,5\}$, with
$I_{3,1} \supset I_{2,1}, I_{2,2}, I_{2,6}$ (see also Examples \ref
{es-forcina} and~\ref{esempio1}).\looseness=1
\end{rem}

Obviously, for a complete subgraph there may be more
than one identifying sequence. The following result shows the
relationship between generalized
identifying sequence for cliques and identifying sequence for
complete subsets.

%pr3 #&#
\begin{prop} \label{prop3} For any complete subgraph $I_{k,r}$ (for
any $k$) of
graph $G^O$ there exists an identifying sequence $\{
I_s\}_{s=0}^{q'+1}$, $I_0=I_{k,r}$, if and only if for
each clique $C_0$ of $G^O$ with $\vert C_0\vert>1$ there exists a
generalized identifying sequence $\{ S_s
\}_{s=0}^{q}$, $S_0=C_0$.
\end{prop}

\begin{pf} It is immediate to see that the existence for a
complete subgraph in $G^O$ of an identifying sequence implies the
condition on the cliques $C$: it is enough for any clique $C$ to
consider the relevant identifying sequence $I^C_{q'+1}$
and then, since $I_{q'+1}^C$ is complete,\vspace*{1pt}
consider again the relevant identifying sequence for $I^C_{q'+1}$ until
the last
term has\vspace*{1pt} cardinality 1. The proof of the inverse implication is
the following. For $S=C_0$, it is trivial. For $S\subset C_0$
consider the following restriction on the sets $S_0,\ldots,S_{q}$
in the generalized identifying sequence for $C_0$: let $I_0=S$ and, for
$i \in
\{1,\ldots,q'+1\}$, let $I_i$ be the subset of nodes $v\in S_i$
such that there exists $j\in I_{i-1}$ with $(j,v)\in\bar E$ and
such that the cardinality of $I_i$ is not greater than $|S|$ (see
Remark \ref{rem-1}). The existence of $I_{q'+1}$ with
$|I_{q'+1}|<|S|$ follows from $|S_{q}|=1$.
\end{pf}

%le1 #&#
\begin{lemma}\label{ex-lem2}
Let $G^K$ be an
undirected graphical model over the binary variables $(A_0,A_1,
\ldots, A_n)$ with $A_0$ latent and with $(0,u)\in E$, for all
$u \in O$. Let $I_{k,r}$ be a complete subgraph of $G^O$ with $k\geq
2$ that admits an identifying sequence $\{I_s\}_0^{q+1}$. Then $D(\beta
)$ contains at least one square
sub-matrix $M_{k,r}$ of order $2(q+1)$ formed by the rows $d_{I_s}$
and $d_{\{V,I_s\}}$, $V\subseteq I_{s+1}$, and by the
columns associated to $\beta_{I_{s}}$ and $\beta_{\{0, I_s\}}$, $s
\in\{0,\ldots,q\}$, that has full rank everywhere in the parameter
space.

Conversely, if $D(\beta)$ is full rank everywhere in the parameter
space, then for any clique $C_0$ of $G^O$ with $|C_0| >1$ there is at
least a generalized identifying sequence.
\end{lemma}

\begin{pf} See Appendix \hyperref[AppA]{A}.
\end{pf}
%

%ex4 #&#
\begin{example} \label{segue}
With reference to Figure {\ref{prima}}, let $I=\{1,2\}$. The
square sub-matrix with rows $d_I$ and $d_{\{4, I\}}$, and columns
$\beta_{I}$ and $\beta_{\{0,I\}}$ is full rank, as the sequence
$I_0=\{1,2\}$, $I_1=\{4\}$ satisfies the assumptions of Lemma
\ref{ex-lem2}. Let $I=\{2,3\}$, the square sub-matrix with rows
$d_I$ and $d_{\{5,I\}}$ and columns $\beta_I$ and
$\beta_{\{0,I\}}$ is also full rank, as the sequence
$I_0=\{2,3\}$, $I_1=\{5\}$ satisfies the assumptions of Lemma
\ref{ex-lem2}. The same holds for $I=\{3,4\}$ and $I=\{4,5\}$,
since for both $I_1=\{1\}$ is the required set.
\end{example}

Suppose that for each fixed order $k$ of interaction, $k\in\{ 2,
\ldots, t\}$, the sets $I_{k,r}$, $r=1,\ldots,s_k$, satisfy the
assumptions of Lemma \ref{ex-lem2}. For each $I_{k,r}$ then there is a
full rank
sub-matrix $M_{k,r}$ of $D(\beta)$ with rows $d_{I_s}$, $d_{\{V,
I_s\}}$, $V \subseteq I_{s+1}$, and columns $\beta_{I_s}$ and
$\beta_{\{0, I_s\}}$, $s \in\{0, \ldots, q\}$. We denote with $P_k$
the matrix formed by all rows of $D(\beta)$ and columns used to
build all the matrices $M_{k,r}, r \in\{1,\ldots,s_k\}$. By
construction, a row, and therefore a column, cannot appear in more
than one $M_{k,r}$. Then, $P_k$ is a sub-matrix of $D(\beta)$ which
is full column rank as it is block-triangular matrix with full-rank
blocks $M_{k,r}$. In fact, the matrix $P_k$ has zero components in
the columns associated to $\beta_{\{I_{k,r'}\}}$ and $\beta_{\{0,
I_{k,r'}\}}$ for $r'\neq r$, so $P_k$ is a lower block-triangular
matrix with blocks full rank everywhere in the parameter space, and
is therefore full rank for all $\beta\in\Omega$. The following
result then holds.

%pr4 #&#
\begin{prop} \label{ex-lem3} Let
$P=[P_2 \mid\ldots\mid P_t]$ be the sub-matrix of $D(\beta)$, with
$P_k$, $k\in\{ 2, \ldots, t\}$, constructed as previously
described. If for any clique $C$ of $G^O$ with $|C_0| >1$ there is a generalized
identifying sequence, then $P$ is full column rank everywhere in
the parameter space.
\end{prop}

\begin{pf} From the fact that the model is graphical, $P$ is
lower block-triangular matrix, as if $\beta_I=0$ then
$\beta_{I'}=0$ for all $I' \supset I$. Then the blocks are full
column rank everywhere in the parameter space by Lemma
\ref{ex-lem2}.
\end{pf}

We can then prove the following theorem.

%th2 #&#
\begin{theo}\label{hip2}
Let $\beta$ be the vector of the parameters of an undirected
graphical model $G^K$ over the binary variables
$(A_0,A_1,\ldots,A_n)$, with $A_0$ latent and $(0,u) \in E$, for
all $u \in O$. A necessary and sufficient condition for $D(\beta)$
to be full rank everywhere in the parameter space is that:
\begin{enumerate}[(ii)]
\item[(i)] $\bar G^O$ contains at least one $m$-clique $C$,
with $m \geq3$;
\item[(ii)] for each clique $C_0$ in
$G^O$ with $|C_0| >1$ there exists a generalized identifying sequence.
\end{enumerate}
\end{theo}

\begin{pf} See Appendix \hyperref[AppA]{A}.
\end{pf}

As already noticed, violation of assumption (i) of Theorem
\ref{hip2} implies that the graph $ G^O$ is composed either by two
and only two complete components that are not connected or by one
connected component. The first case has been discussed in Section
\ref{s3} and leads to models that are not even generically identified.
The second case leads to models that are locally identified almost
everywhere in $\Omega$. The subset where identification breaks
down is derived in Appendix \hyperref[AppA]{A}.

Violation of assumption (ii) of Theorem \ref{hip2} implies that there
is a
subspace of null measure in which $D(\beta)$ is not full rank, which can
be so determined. If there is a clique having no generalized
identifying sequence, there is (at least) a complete set $I_0$ in
$G^O$ having no complete set $I_1$ in $G^O$ containing nodes that
are connected in $\bar G^O$ to a node of $I_0$. Then, we need to
find the set $bd_{\bar G^O}(I_0)$ of nodes adjacent to at least a
node in $I_0$ in the complementary graph $\bar G^O$. In this set,
find all $V_0$ subsets that are complete in $G^O$. The expression
of the subspace may be derived by equation (\ref{con}) in Appendix \hyperref[AppA]{A}.
This is:
%e4 #&#
\begin{equation}\label{con5}
\beta_{\{0,V_0\}}+\sum_{I\subseteq
I_0} \delta(V_0,I)\beta_{\{0,I,V_0\}}=0  \qquad \mbox{for any } V_0
\subseteq bd_{\bar G^O}(I_0),
\end{equation}
where $\delta(V_0,I)=1$ if $\{V_0,I\}$ is complete in $G^O$ and 0
otherwise. Note that the sets that have a non-zero contribution to $\sum
_{I\subseteq I_0} \delta(V_0,I)\beta_{\{0,I,V_0\}}$
are necessarily subsets of $I_0^{V_0}$ sets.

%ex5 #&#
\begin{example} \label{ex-es3}
Let the cliques in the graph $G^O$ be the following $C_1=\{
1,4,7,9\}, C_2=\{ 1,4,6,9\},\allowbreak C_3=\{ 1,4,6,8\}, C_4=\{ 2,4,7,9\},
C_5=\{ 2,4,6,9\}, C_6=\{ 2,4,6,8\}, C_7=\{ 1,5,\break7,9\}, C_8=\{
2,5,7,9\}, C_9=\{ 3,5,8\}, C_{10}= \{3,6,8\}, C_{11}=\{ 1,5,8\},
C_{12}=\{ 2,5,8\}, C_{13}=\{3,5,7\}$. In Figure {\ref{terza}}(a)
and (b) the corresponding graphs $G^O$ and $\bar G^O$ are
represented. We can verify from the graph $\bar G^O$ that the
assumptions of the Theorem \ref{hip2} hold. For example, for the
clique
$C_{11}=C_0=\{ 1, 5,8\}$ we have the generalized identifying
sequence: $S_1=\{ 2, 4,9\}, S_2=\{ 3\}$. By considering
$C_3=S_0=\{ 1, 4, 6,8\}$ we have the generalized identifying
sequence $S_1=\{ 3, 7\}$, $S_2=\{ 4,6\}$ and $S_3=\{ 5\}$. The
corresponding graphical model is therefore locally identified everywhere
in the parameter space.
\end{example}

% Figure \ref{quarta} the
%is generically identified, as condition {\em(i)} of Theorem
%does not hold for $C_0=\{4,5,8,9\}$ as the boundary in $\bar G^O$
%is $\{6,7\}$ and $(6,7) \in\bar E$. The subspace where
%identifiability breaks down
%is so constructed. See the file Antonio

%{\bf Anche questo esempio \ref{ex-es5} lo toglierei e la figura
%associata \ref{quinta}}
%
%$(ii)$ of Theorem \ref{hip2} does not hold. As a matter of fact,
%for $\{1, 3,7\}$ and $\{ 3, 4,6,7\}$: in fact, in $\bar G^O$ node
%$3$ is connected only to node $5$ and node $7$ is connected only
%to node $2$, with $(2,5)\in\bar E$. By adding the edge $(3,7)$ in
%the missing graph $\bar G^O$, we get a local identified model.

%ex6 #&#
\begin{example}\label{es6} The model associated to the graphs in Figure \ref{sesta}
satisfies condition (i) of Theorem
\ref{hip2}. However, condition (ii) does not hold for
$\{1,2,3,4\}, \{4,5\},\{4,6\}$.

For $C_0=\{1,2,3,4\}$ we have
$bd_{\bar G^O}(C_0)=\{5,6\}$, which is complete in $\bar G^O$. Then, the
complete sets $V_0$ are $\{5\}$ and $\{6\}$. From (\ref{con5}),
$V_0=\{5\}$ ($V_0=\{6\}$) gives rise to the first (second) equation
of the system below.

%f4 #&#
\begin{figure}

\includegraphics{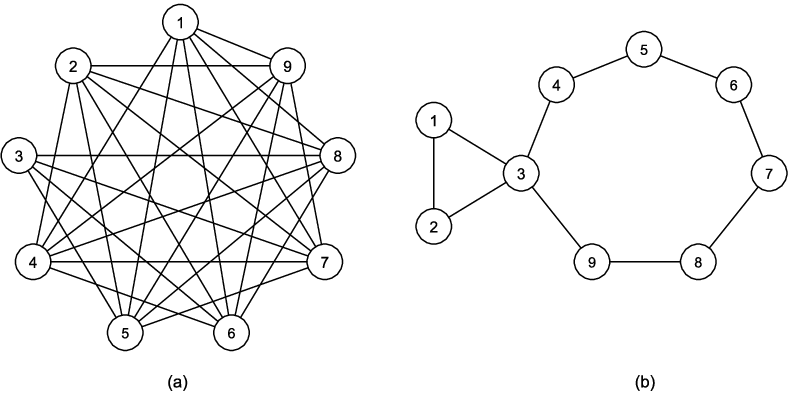}

\caption{The graph (a) $G^O$ and (b) its complementary graph
$\bar G^O$ of the model in Example \protect\ref{ex-es3}.} \label{terza}
\end{figure}

%f5 #&#
\begin{figure}[b]

\includegraphics{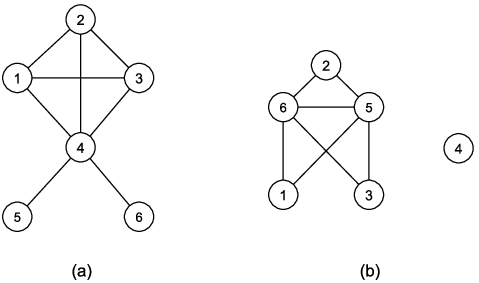}

\caption{The graphs (a) $G^O$ and (b) $\bar{G}^O$ of Example
\protect\ref{es6}.} \label{sesta}
\end{figure}

For
$C_0=\{4,5\}$, the $bd_{\bar G^O}(C_0)=\{1,2,3,6\}$. The sets $V_0$
are all possible complete subsets of $\{1,2,3\}$ and $\{6\}$. From
(\ref{con5}), the equations of the system below are formed, with the
exclusion of the first one. Analogously, for $C_0=\{4,6\}$ the
equations of the system below are formed, with
the exclusion of the second one. So we have:
\[
\cases{
\beta_{\{0,5\}}+\beta_{\{0,4,5\}} =0,\cr
\beta_{\{0,6\}}+\beta_{\{0,4,6\}} =0,\cr
\beta_{\{0,1\}}+\beta_{\{0,1,4\}} =0,\cr
\beta_{\{0,2\}}+\beta_{\{0,2,4\}} =0,\cr
\beta_{\{0,3\}}+\beta_{\{0,3,4\}} =0,\cr
\beta_{\{0,1,2\}}+\beta_{\{0,1,2,4\}} =0,\cr
\beta_{\{0,1,3\}}+\beta_{\{0,1,3,4\}} =0,\cr
\beta_{\{0,2,3\}}+\beta_{\{0,2,3,4\}} =0,\cr
\beta_{\{0,1,2,3\}}+\beta_{\{0,1,2,3,4\}} =0.}
\]
The rank of $D(\beta)$ is equal 40 everywhere except in the subspace
above.
Notice that clique $C_0=\{1,2,3,4\}$ contains the following 7 complete subsets
with cardinality greater than 1 having no identifying sequence: $\{
1,2,4\}$,
$\{1,3,4\}$, $\{2,3,4\}$, $\{1,4\}$, $\{2,4\}$, $\{3,4\}$. For all
these sets, $bd_{\bar G^O}(I_0)=\{5,6\}$ and from (\ref{con5}) the
first two equations of the system above are formed. From these derivations,
we can see that some intermediate situations can occur: the rank of
$D(\beta)$ degenerates of 8 in the subspace formed by the
first (second) equation only while it degenerates of 2 in the subspace
formed by the last seven
equations only.
While in the subspace given from all the above equations, $D(\beta)$
degenerates to 30 due to the 9 complete subsets with no identifying
sequence (i.e., the 7 aforementioned complete subsets of $\{1,2,3,4\}$
plus $\{4,5\}$ and $\{4,6\}$) and to the fact that the node $4$ is not
connected to the other nodes in $\bar G^O$ (see (\ref{pippo}) of
Proposition \ref{pippo1}).
\end{example}

We now extend the condition for local identification to more general
models with observable random variables $v \in O$ with a finite number of
levels $l_v$.

%th3 #&#
\begin{theo} \label{teo-ex1}
Let $\beta$ be the vector of parameters of an undirected graphical
model $G^K$ over the discrete
variables $(A_0,A_1,\ldots,A_n)$, with $A_0$ latent binary
variable and $(0,u)\in E$, for all $u \in O$. A necessary and
sufficient condition for $D(\beta)$ to be full rank everywhere in
the parameter space is that:
\begin{enumerate}[(ii)]
\item[(i)] $\bar G^O$ contains at least one $m$-clique $C$,
with $m \geq3$;
\item[(ii)] for each clique $C_0$ in
$G^O$ with $|C_0| >1$ there exists a generalized identifying sequence.
\end{enumerate}
\end{theo}

\begin{pf} See Appendix \hyperref[AppA]{A}.
\end{pf}

All models that are locally identified for the binary case are also
identified for the more general case,
provided that the latent variable is binary. Note that, for models
that are locally identified everywhere except in a subspace of null
measure, the equation of the subspace can be found by making
repeated use of equation (\ref{con5}), after noting that the
parameters expressing the interaction terms of a subset $I$ are as
many as the product of the levels $\prod_v(l_v-1)$, $v \in I$.

Note that, for the particular case of a binary hidden variable,
Theorem \ref{teo-ex1} extends the class of (generically) identified models
according to Allman \textit{et al.} \cite{Allmanetal}, as their
identification criteria allows for conditional independence
between blocks of observable variables given the latent one only,
and therefore excludes models with $G^O$ connected. Note further
that Theorem \ref{teo-ex1} implies that only the models with
connected complementary graph can be identifiable. This contrasts
with the condition of globally identifiability in graphical
Gaussian models given in \cite{Stanghellini,Vicard}. The two
conditions coincide only in the case with $n=3 $ or $n=4$. In this
second case, an identified model (under both the discrete and
Gaussian distribution) has conditional independence graph as in
Figure \ref{esstanghe}(a).

%s5 #&#
\section{Concluding remarks}\label{s5}

One of the issues in estimating graphical models with latent
variables concerns identifiability. In this paper, a
characterization of locally identified undirected discrete
graphical models with one hidden binary node has been presented,
through a necessary and sufficient condition which can be checked
from the associated concentration graph. Investigation on the
consequences of violation of the given condition led to
distinguish between models that are locally identified everywhere
but in a subspace of null measure and models that are not locally
identified. In the first case, the derivations allow to determine
the subspace of null measure where identifiability fails.

Issues of identification of all models that are obtainable as a one to one
reparametrization of the discrete undirected graphical model can
be addressed using the results here presented. We also conjecture that
results on block-triangularity of the matrix $D(\beta)$ can be
extended to deal with models with one discrete latent node with more
than two levels. The derivations in this paper
also pave the way to graphical models with more than one hidden
variable as well as directed acyclic graphs.

%% The Appendices part is started with the command \appendix;
%% appendix sections are then done as normal sections

%apA #&#
\begin{appendix}

\setcounter{equation}{0}

\section{\texorpdfstring{Proofs of derivations in Section \protect\ref{s4}}{Proofs of derivations in Section 4}}\label{AppA}
\subsection*{Algorithm for reordering $D(\beta)$}

Let $U\subseteq\bar C$ be
the set of unordered nodes. Given $C$, for any $v$ node in $\bar C$,
let $\pi_v$ be (one of) the shortest paths
connecting $v$ to a node in $C$ and let $\lambda_v$ be its
length. This path exists whenever the graph $\bar G^O$ is connected.
Let $a_i$ be (one of) the farthest node among those in
$U\subseteq\bar C$ such that $\lambda_{a_i}=\max_{v\in U} \lambda
_v$. Let $W_i$ be the
ordered set of nodes in the path $\pi_i$ in the direction emanating
from $C$ to $a_i$. (The path $\pi_i$ may contain
nodes which do not belong to $U$.) Denote with $b_i$ the last
node of $W_i$ belonging either to $C$ or to $\bar C\setminus U$.

\begin{enumerate}[Step 3.]
\item[Step 1.] $U\leftarrow{\bar C}$, $T\leftarrow C$.
\item[Step 2.] Check if $U$ is empty, in this case $\bar C$ is ordered;
otherwise search for the $a_i$ node, with the corresponding $W_i$ and
$b_i$.\eject
\item[Step 3.] Let $J_i$ be the ordered set obtained from $W_i$ by
deleting the elements before $b_i$ and~$b_i$.
\item[Step 4.] If $b_i$ is in $C$, then append $J_i$ to $T$ as the last
group of elements (so $T\leftarrow\{T, J_i\})$; otherwise, if $b_i$
is in $T\setminus C$ order $J_i$ just after $b_i$ in $T$ (so
$T\leftarrow
\{C, \ldots, b_i,J_i, \ldots\}$);
\item[] let $U\leftarrow U\setminus J_i$; go to Step 2.
\end{enumerate}

\subsection*{Proof of Lemma \protect\ref{ex-lem2}}
We prove the sufficiency first. Consider
all the sub-matrices of $M_{k,r}$. Observe that a row, and therefore
a column, cannot be chosen twice in a $M_{k,r}$ matrix, as $I_s\neq
I_{s^{\prime}}$ (see Remark~\ref{rem3}). By ordering the rows and columns
according to the
sequence of $\{I_s\}_0^{q+1}$, the matrix $M_{k,r}$ is seen to be lower
block triangular. The blocks are $N_0,\ldots,N_q$ where $N_s$ is
formed by the rows $d_{I_s}$ and $d_{\{V,I_s\}}$ with $V\subset
I_{s+1}$ (from Remark \ref{rem0} the intersection $I_s$ and $I_{s+1}$
is empty) by the columns
associated to $\beta_{I_{s}}$ and $\beta_{\{0, I_s\}}$. Therefore,
$N_s$ is as\vspace*{1pt} in~(\ref{primosistema}).

Then,
$\rank(M_{k,r})=\sum_{s=0}^q\rank(N_s)$ and is full if
and only if the blocks are full rank, that is if the rank of each
block is equal to 2.

Suppose that there is no index $s$ such that $N_s$ has full rank, that
is, there is no
$V\subseteq I_{s+1}$ generating a sub-block of $N_s$ with rank equal to 2.
Then, from (\ref{primosistema})
\[
\sum_{I\subseteq\{I_{s} \cup V\}, I \not\subseteq I_s }
\delta(I)\beta_{\{0 ,I\}}=0 \qquad\mbox{for all }
V \subseteq I_{s+1}.
\]
From the fact that the
model is graphical, we obtain:
%
%e1 #&#
\begin{equation}\label{syst}
\sum_{I \subseteq I^V_s}\beta_{\{0,V, I\}}=0 \qquad
\mbox{for all } V\subseteq I_{s+1},
\end{equation}
where for $V=I_{s+1}$ one has $I^{s+1}_s=\emptyset$. This implies that
$\beta_{
\{0,V\}}=0$, which contradicts the assumptions since
$I_{s+1}$ is a complete subgraph of $G^O$. Therefore, for each $s$
there exists
a full rank block $N_s$ and the square sub-matrix
$M_{k,r}$ is full rank everywhere in the parameter space.

We now
prove the necessity. Since $D(\beta)$ is full rank everywhere, the
sub-matrix of $D(\beta)$ formed by all rows of $D(\beta)$ and by
the columns $\beta_{I_{k,r}}, \beta_{\{0,I_{k,r}\}}$ is full
column rank for all $\beta\in\Omega$. Going by contradiction,
suppose that there is a clique $C$ in $G^O$ admitting no
generalized identifying sequence. Then, from Proposition
\ref{prop3} there is a $I_{k,r}=I_0$ such that there is no
identifying sequence. Then, we can suppose without loss of
generality that there is no complete subgraph $I_1$ in $G^O$ such
that for each $i\in I_0$ there is $j\in I_1$ with $(i,j)\notin
E$. Select the sub-matrix $C_{k,r}$ formed by the columns
$\beta_{I_0}$, $\beta_{\{0, I_0\}}$ and all the rows such that these
two columns have non-zero components, that is select all rows
$d_V$, $V\supseteq I_0$. (Note that in all other rows the two
elements are both 0.) Denote with $\Omega_{k,r}\subset\Omega$ the
following subspace:
%
%e2 #&#
\begin{equation}\label{con}
\beta_{\{0, V_0\}}+\sum_{I\subseteq
I_{0}^{V_0}}\beta_{\{0,I,V_0\}} =0,
\end{equation}
where $V_0$ is any complete subgraph in $G^O$ such that
for each $j\in V_0$ there is at least a $i\in I_0$ with $(i,j) \in
E$. Violation of assumption (a) of Definition \ref{id-seq}
implies that $I_{0}^{V_0} \neq\emptyset$. Then, it is easy to
verify that for $\beta\in\Omega_{k,r}$ as defined by (\ref{con})
the columns of $C_{k,r}$ are linearly dependent. As a matter of
fact, every $2 \times2$ sub-matrix formed by any two rows of
$C_{k,r}$ has the form of (\ref{primosistema}) with $b'=0$.
This contradicts the assumption that $D(\beta)$ is
full rank everywhere.

Suppose now the violation of assumption (b) of Definition
\ref{id-seq}, that is, that there exists a $I_0=I_{k,r}$ such that
there is no sequence for $I_{0}$ such that $|I_{q+1}|<k$. We
can find a $ 2 \times2$ full rank submatrix of $D(\beta)$ with
columns associated to $\beta_{I_s}$ and $\beta_{\{0, I_s\}}$,
$s\in\{0, \ldots, q\}$. From the previous derivations, we should
consider the rows associated to $I_s$ and $V_S=\{I_s, I_{s+1}\}$
(otherwise $D(\beta)$ is not full rank for $\beta\in
\Omega_{k,s}$ as defined by (\ref{con})). But, as there is no
$I_{q+1}$ such that $|I_{q+1}|<k$, $I_{q+1}$ coincides with some
$I_s$ in the sequence. Therefore, we cannot find the required $2
\times2$ sub-matrix with full rank.

\subsection*{Proof of Theorem \protect\ref{hip2}}
We prove the sufficiency first. Let
$D_C$ be the
sub-matrix of $D(\beta)$ with rows corresponding to the cells
with values zeros for all variables not in $C$, and columns $\mu,
\beta_i,\beta_{\{0, i\}}$, $i \in C$. By (i) the graph $G^C$
corresponds to a binary latent class model and so by Proposition
\ref{prop1}, $D_C$ is full column rank. Let $D_{\bar C}$ be the
sub-matrix of $D(\beta)$ having rows $d_i$, $d_{\{i, j\}}$ and
columns $\beta_i,\beta_{\{0, i\}}$, $i \in\bar C$ and $j$ such
that $(i,j)\in\bar E$ ($j$ could belong to $C$). From (ii) and
Proposition \ref{prop3}, it follows that for any complete subgraph
in $G^O$ there is an identifying sequence. Then, from Lemma
\ref{ex-lem2}, $D_{\bar C}$ is full column rank. The matrix
$D(\beta)$ can be so written:
\[
D(\beta)=\left[
\matrix{
D_C & 0 & 0\cr
B_1 & D_{\bar C} & 0 \cr
B_2 & B_3 & P\cr}
\right],
\]
where $B_1$, $B_2$ and $B_3$ are non-zero matrix (we omit the
dimension for brevity), while $P$ is as in Proposition
\ref{ex-lem3}. Therefore, $D(\beta)$ is full rank everywhere.

To prove the necessity, it is enough to note that $D(\beta)$ is
full rank only if the following matrices $D_C$, $D_{\bar C}$ and
$P$ are full rank. If $D_C$ is full rank, then by Proposition
\ref{prop1}, condition (i) holds. From Lemma \ref{ex-lem2},
$D_{\bar C}$ and $P$ full rank imply that for any clique of $G^O$
there is a generalized identifying sequence, and so by Proposition
\ref{prop3}, condition (ii) holds.

\subsection*{Subset where identifiability breaks down in models
with no $m$-clique in $\bar G^O$, $m \geq3$, and $G^O$ is
connected}
 If there is no $m$-clique, $m \geq3$, then for any
triple of nodes $i_1,i_2,i_3$ there is at least an edge between
two of them missing in $\bar G^O$. Consider the sub-matrix
$D_3(\beta)$ of $D(\beta)$ related to the rows $\{(0,0,0),
(0,1,0), (1,1,0), (1,0,0),
(0,0,1),
(0,1,1), (1,1,1), (1,0,1)\}$ and
columns $\mu,\beta_0,\beta_{i_r},\beta_{\{0,i_r\}}$, $r=1,\ldots,3$.
Then the matrix $D_3(\beta)$ has the following structure:
\[
D_3(\beta)=\left[
\matrix{
D_2(\beta) & 0_{4 \times2} \cr
D_2(\beta) & P_{4 \times2} \cr
}\right],
\]
where $0_{4 \times2}$ is a zero sub-matrix and
\[
D_2(\beta)=\left[
\matrix{
a_{11}& a_{12} & 0 & 0 & 0 & 0 \cr
a_{21}& a_{22} & 0 & 0 & a_{21} & a_{22} \cr
a_{31}& a_{32} &a_{31}& a_{32} & a_{31}& a_{32} \cr
a_{41} & a_{42} & a_{41} & a_{42} & 0 & 0 \cr
}\right].
\]
Note that the generic elements of the matrix
$D_2(\beta)$ are
\begin{eqnarray*}
a_{i1}&=&\mathrm{e}^{\mu+\sum_{I\subseteq{I_i}} \delta(I) \beta_I}
\bigl(1+\mathrm{e}^{\beta_0+\sum_{I\subseteq{I_i}} \delta(I)
\beta_{\{0,I\}}}\bigr),\\
a_{i2}&=&\mathrm{e}^{\mu+\beta_0+\sum_{I\subseteq{I_i}} \delta(I) \beta_{\{
0,I\}}}
\end{eqnarray*}
with $I_i$ the set of random variables taking value 1 in
row $i$ and $\delta(I)=1$ if $I$ is complete in $G^O$.
The matrix $D_2(\beta)$ is not
full rank in the subspace of $\Omega$ where all the $4 \times4$
square sub-matrix
of $D_2(\beta)$ are not full rank. Analogously, the matrix $P_{4
\times2}$ is not full rank for $\beta$ in $\sum_{I\subseteq{I_i}, I
\not\subseteq I_j} \delta(I) \beta_{\{0,I\}}=0$ for all $i,j \in{\{
5,\ldots,8\}}$, with $j>i$.

\subsection*{Proof of Theorem \protect\ref{teo-ex1}}
First, assume that all the
variables are
binary except the $A_1$ variable which has three levels. Partition
$\beta$ into three subsets $\beta^a =\{\mu$, $\beta_0\}$, $\beta^b$
corresponding to the non-zero interaction terms of any order for
value in $\{0,1\}$ of the observable random variables and $\beta^c$
containing all other parameters. After ordering in a way such that
the $A_1$ variable is running the slowest, the $D(\beta)$ matrix has
the following structure:
\[
D(\beta)=\left[
\matrix{
 D(\beta^a)  &D(\beta^b) & 0_{2^n\times
|\beta^c|} \cr
 D^{*}(\beta^a)& 0_{2^{(n-1)} \times|\beta^b|}  &
D^{*}(\beta^c)
}\right],
\]
where $[ D(\beta^a) \mid D(\beta^b) ]$ is
the sub-matrix of the derivatives of $\beta^a$ and $\beta^b$. It
has full rank if conditions (i) and (ii) of Theorem \ref{hip2}
hold. Note that by construction, $D^{*}(\beta^c)$ has a similar
structure of the sub-matrix of $D(\beta^b)$ formed by the last
$2^{(n-1)}$ rows and all columns. Therefore, $D^{*}(\beta^c)$ is
full rank if conditions (i) and (ii) of Theorem \ref{hip2}
hold.

To see the
necessity note that $D(\beta^b)$ is full rank only if Theorem \ref{hip2}
is verified. Proof of the theorem for $A_1$ having $l_v$ levels
follows straightforwardly. By a similar argument, extension to a
generic number of levels of the $A_i$ variables, $i \in O$, follows.
%

%All the $4 \times4$ square sub-matrix of $D_2(\beta)$
%are not full rank, so also $D_2(\beta)$ and $D_3(\beta)$ are not
%full rank. \rosso{By solving such sub-matrices the subspace where
%identifiability breaks down can be determined}.

\section{\texorpdfstring{Proof of Theorem \protect\ref{teo-ex2}}{Proof of Theorem 1}}\label{AppB}

Note that $T_1$ is the set
of observable variables such that $(i,O) \notin E$. We first focus
on models with only binary variables. Let $T_2\subseteq S\setminus
\{ 0\}$ be the set of observable variables such that $(i,j) \in E$,
$i\in T_1$, $j \in T_2$. If $T_1$ or $T_2$ is empty the proof is
trivial.

To start
with, we assume $|T_1|=1$. Partition $\beta$ into the subsets
$\beta^d$ containing all the non-zero interaction terms among
the variables in $S$ and $\beta^e$ containing all the other
elements. The non-zero interaction terms among the latent variable
and the observable random variables are in $\beta^d$. The matrix
$D(\beta)$ has the following structure:
\[
D(\beta)=\left[
\matrix{
 D(\beta^d) &  0_{2^{|S|-1} \times|\beta^e|} \vspace*{2pt}\cr
 F &  D(\beta^e)
}\right],
\]
$D(\beta^d)$ and $D(\beta^e)$ are the derivative
sub-matrices for the corresponding elements.
%; $F$ is a sub-matrix with the same number of rows of $D(\beta^d)$.
The sub-matrix
$D(\beta^e)$ is full rank because it corresponds to the rank of the
design matrix of the model for $T_1 \cup T_2$. The conclusion follows
easily from the
block-diagonality of the matrix and from the fact that by Theorem
\ref{teo-ex1} $D(\beta^d)$ has full rank if and only if (i) and
(ii) hold. Extension to a generic number of
variables in $T_1$ follows after noting that the matrix $D(\beta)$
is so built:
\[
D(\beta)=\left[
\matrix{
D(\beta^d) &  0_{2^{|S|-1} \times|\beta^e|}\vspace*{2pt}\cr
 F^*  & D(\beta^e)
}\right],
\]
where $D(\beta^e)$ is the derivative sub-matrix for the
vector $\beta^e$ defined as in the previous step. $D(\beta^d)$ is
the derivative sub-matrix for the vector $\beta^d =\beta\setminus
\beta^e$; $F^*$ is a sub-matrix with the same number of rows as
$D(\beta^e)$. The same considerations as in the previous case
hold. Extension to a generic number of levels of the $A_i$
variables, $i \in O$, follows by induction, as done in the proof
of Theorem~\ref{teo-ex1}.
\end{appendix}

\section*{Acknowledgements}

We are grateful to Antonio Forcina for writing a set of Matlab
routines by which one can easily check the main results of the paper, as
well as for stimulating discussions and comments. We also thank the
referees for their very detailed
and constructive criticism.

%suskaldyti doi

% imsref loaded by aiste.veprauskaite, 2012-06-14 12:22:00
% imsref loaded by aiste.veprauskaite, 2012-06-14 12:25:39

\printhistory

\end{document}